\documentstyle[epsfig,longtable]{aipproc}

\begin{document}


\title{Stereo imaging of the VHE $\gamma$-rays with HEGRA \& H.E.S.S.}

\author{Alexander K  Konopelko}
\address{Max-Planck-Institut f\"ur Kernphysik,\\ 
D-69117 Heidelberg, Postfach 10 39 80, Germany\\
e-mail: alexander.konopelko@mpi-hd.mpg.de}

\maketitle

\begin{abstract}
At present, the ground-based astronomy of very high energy (VHE)
(E$>$100~GeV) $\gamma$-rays experiences the challenging transition
phase caused by the substantial upgrading of its observational
instrumentation. Recently the HEGRA collaboration has evidently
demonstrated the power of arrays of imaging atmospheric \v{C}erenkov
telescopes and initiated a number of complementary projects,
such as H.E.S.S., CANGAROO IV, VERITAS. The common philosophy of those
projects is based on the stereoscopic observations of air showers with
a number of imaging telescopes simultaneously. Such observations allow
to reduce substantially the effective energy threshold of detected
$\gamma$-ray showers; to improve the angular and energy resolution
for individual $\gamma$-rays; to gain in suppression of the cosmic 
ray background. In my talk I have summarized the general advantages 
and achieved sensitivity of the stereoscopic observations using the 
HEGRA data taken with an array of 5 imaging telescopes having the 
energy threshold of 500 GeV. These results completed with the relevant
simulations may be extrapolated on a firm ground towards the future
H.E.S.S. (phase I) array of four 12 m imaging telescopes with the 
estimated energy threshold below 100 GeV. Here I discuss in short 
the anticipated performance of such an array. 
\end{abstract}

\section*{New Physics Frontiers}

The recent achievements of the VHE $\gamma$-ray astronomy 
offer a great number of diverse and
fundamental physics results which make further developments in
the field highly physically motivated. Future $\gamma$-ray
observations will deliver exhaustive knowledge to understanding the
origin of the cosmic rays by observing the supernovae remnants 
(SNRs) and the physics of the relativistic jets in the active
galactic nuclei (AGN), the measurements of the infrared (IR)
absorption of $\gamma$-rays propagating in the interstellar medium,
the necessary data to prove the mechanisms of a particle
acceleration in the pulsars and plerions, clues to the enigma
of the gamma-ray bursts {\it etc}. At the same time the significant
advancement towards the new physics frontiers will be only possible
after a significant improvement of the observational instrumentation.
The efficient search and detailed study of the sites of the particle
acceleration up to multi-TeV energies, which may brighten up in $\gamma$-rays, 
need lowering farther the energy threshold of the instruments down to
50-100~GeV, ability to localize the $\gamma$-ray emission with
accuracy better than 1~arcmin, large telescopes' field of view for the 
comprehensive studies of the extended $\gamma$-ray emission, large
dynamic energy range (50~GeV - 50~TeV) and good energy resolution
($\leq 10$\%) for precise measurements of the $\gamma$-ray energy
spectra, larger $\gamma$-ray statistics at the time scale less than
1~hr for efficient search for the episodic or highly variable sources 
{\it etc}. The forthcoming {\it stereoscopic} arrays of the imaging
air \v{C}erenkov telescopes, like H.E.S.S. (High Energy Stereoscopic
System), will exactly meet all those requirements. One can reach this
conclusion using both the observational experience with the currently
operating HEGRA (High Energy Gamma Ray Astronomy) instrument
(hereafter by HEGRA I denote the system of 5 IACTs operated
by the HEGRA collaboration) as well as the specific Monte Carlo
simulations. All that allows to predict reliably the performance of
the H.E.S.S. which is briefly summarized below with the emphasize on
the further advancements in the physics analysis of the observational
data. 

\section*{Hardware parameters of IACTs arrays}

Although a good number of hardware parameters is different for
H.E.S.S. as compared with HEGRA one can pick out several of those
which are of the utmost importance. They are -- {\it (i)} the larger
diameter of the reflector; {\it (ii)} better optics
making use of a stable dish and better mirrors alignment; 
{\it (iii)} smaller angular size of the pixels.       

\begin{table}[htbp]
\caption{The hardware parameters of HEGRA and H.E.S.S.}
\begin{tabular}{lll}
                     & HEGRA & H.E.S.S.       \\ \hline
$\bullet$ Reflector diameter   & 3 m   & 13 m \\
$\bullet$ Davies-Cotton design & Yes   & Yes  \\
$\bullet$ Focal length         & 5 m   & 15 m \\
$\bullet$ Number of channels   & 271   & 960  \\
$\bullet$ Ang. pixel size, deg & 0.25  & 0.15 \\
$\bullet$ Field of view, deg   & 4.3   & 5.0  \\
$\bullet$ Signal sampling      & 120 MHz FADCs & 1 GHz ARS \\
$\bullet$ Dynamic range        & 1-500 ph.e. & LG: 1-100 ph.e./HG:16-1600 ph.e. \\
$\bullet$ Trigger              & 2NN/271 & 4/960(sect.) \\ \hline
\end{tabular}
\end{table}

Due to the large reflector and better efficiency of the
photon-to-photoelectron conversion the H.E.S.S. telescopes will detect 
by a factor of $\simeq$10 more \v{C}erenkov light from the same
shower comparing with HEGRA. Such an advantage in the effective
reflecting area of the telescopes enables to reduce the energy
threshold of the observed air showers from 500~GeV, as for the HEGRA,
to the energy threshold within 50-100~GeV for H.E.S.S. 
For such low energy threshold the
observed air showers will place effectively at the higher height above the
observation level and in addition will have more narrow lateral
divergence of the charges particles emitting \v{C}erenkov light. All
that gives effect to the angular size of the \v{C}erenkov light images 
which become relatively small. In order to measure effectively the
angular shape of such images the future telescopes need to have a
better optics and fainter camera pixellation. Note that the high
quality stereoscopic observations with the arrays like H.E.S.S. need
relatively large separation of the telescopes of about $\simeq$120~m. 
\begin{figure}[t]
\begin{center}
\includegraphics[width=0.48\linewidth]{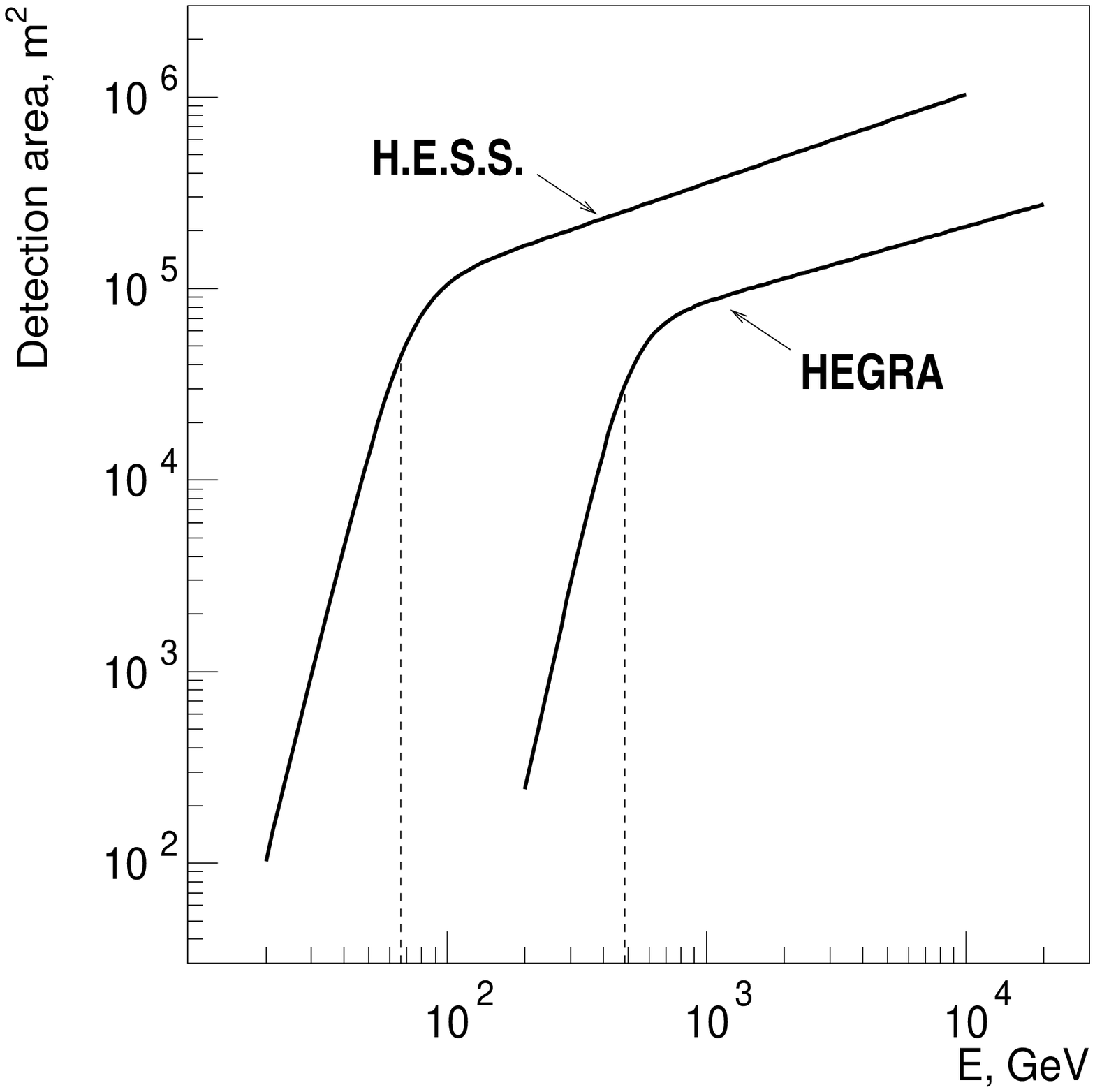}
\includegraphics[width=0.49\linewidth]{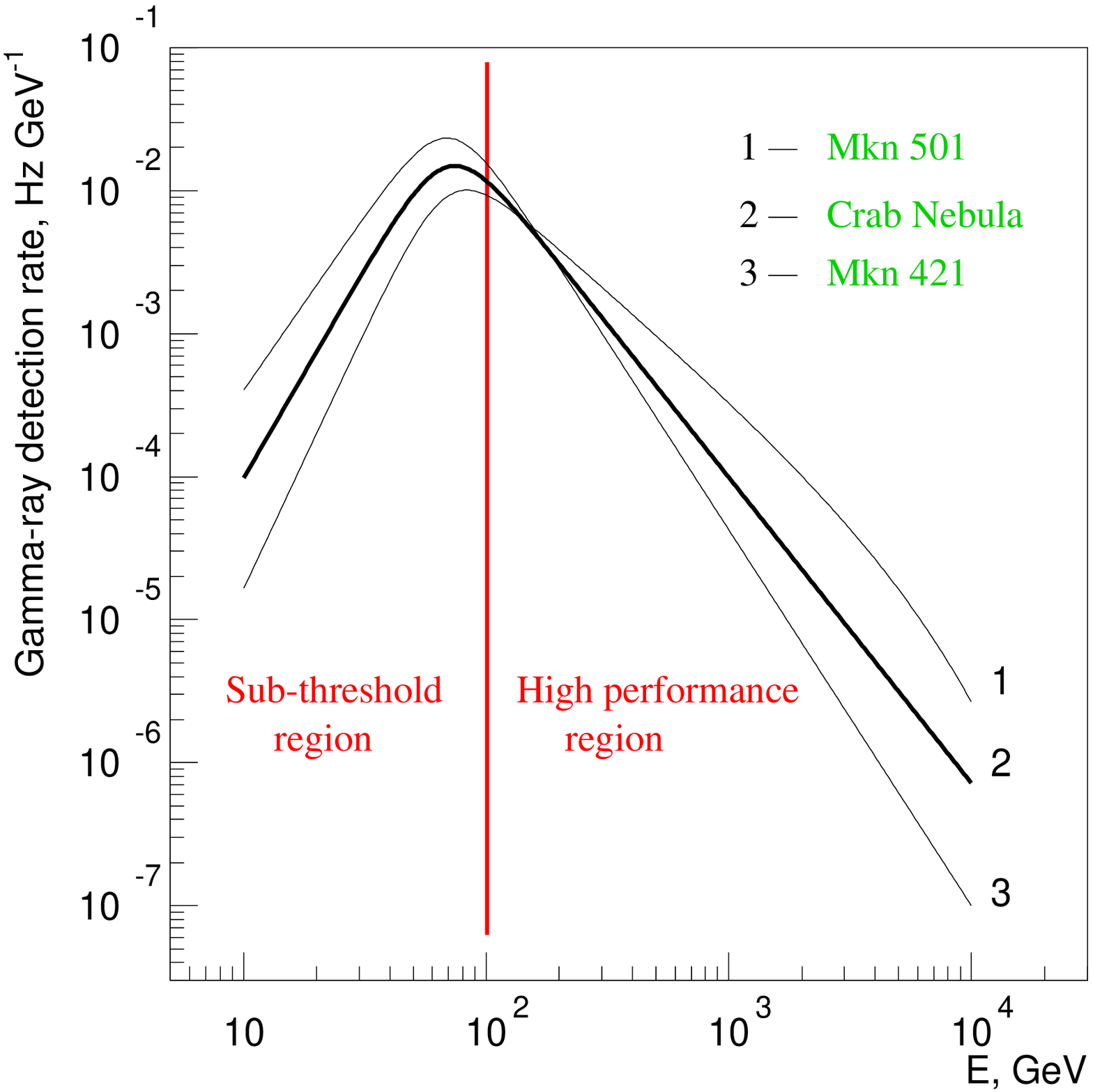}
\end{center}
\caption{The effective detection areas for HEGRA (Trigger: 2/5,
  2NN/271$>$8~ph.e.) and H.E.S.S. (phase I: 4 CTs) (Trigger: 2/4,
  4/960(sect.)$>$4~ph.e.) (left panel). The differential detection rate for 
  H.E.S.S. assuming the initial $\gamma$-ray energy spectrum as for
  Crab Nebula, Mkn~421, Mkn~501, measured with HEGRA IACTs system
  (right panel).} 
\end{figure}    

\section*{Energy Threshold \& Rates}

The effective collection areas of the $\gamma$-ray air showers for the 
arrays of IACTs may be well represented by the following fit:
\begin{equation}
A(E)=a_0 E^{a_1}/(1+(E/a_2)^{a_3})
\end{equation}
where $a_2$-parameter defines the energy threshold, which is of
500~GeV for HEGRA and $\simeq$80~GeV for H.E.S.S. (see Figure~1). 
HEGRA counts of about 120 $\gamma$-ray events per one hr of the Crab
Nebula observations close to the zenith and before applying the
analysis cuts. Assuming the energy spectrum of the Crab Nebula as
measured by HEGRA  
\begin{equation}
dJ_\gamma(E)/dE = 2.8\cdot 10^{-11}E^{-2.6} \rm [photon \, m^{-2} s^{-1}
TeV^{-1}]
\end{equation} 
\begin{figure}[t]
\begin{center}
\includegraphics[width=0.48\linewidth]{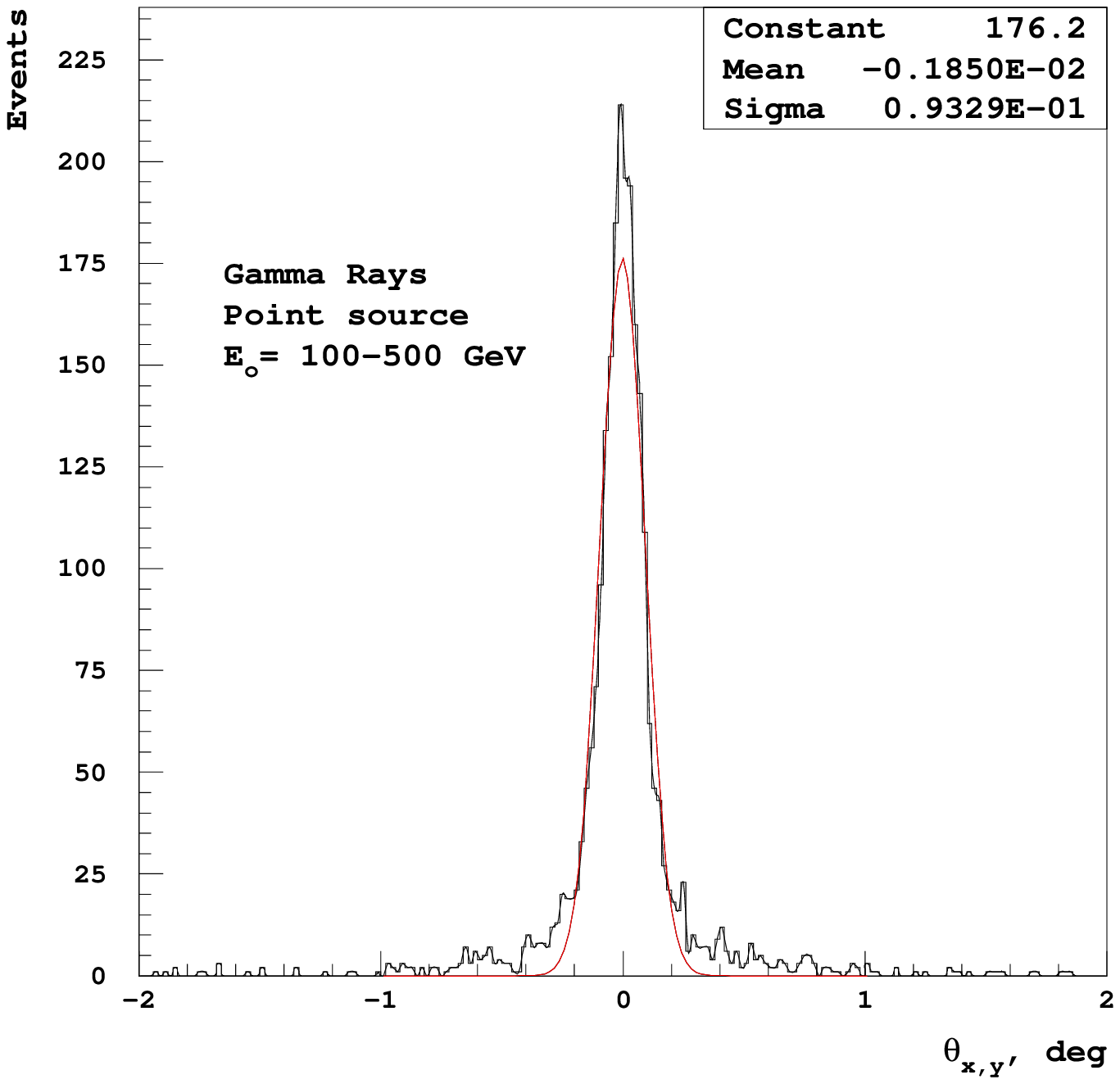}
\includegraphics[width=0.49\linewidth]{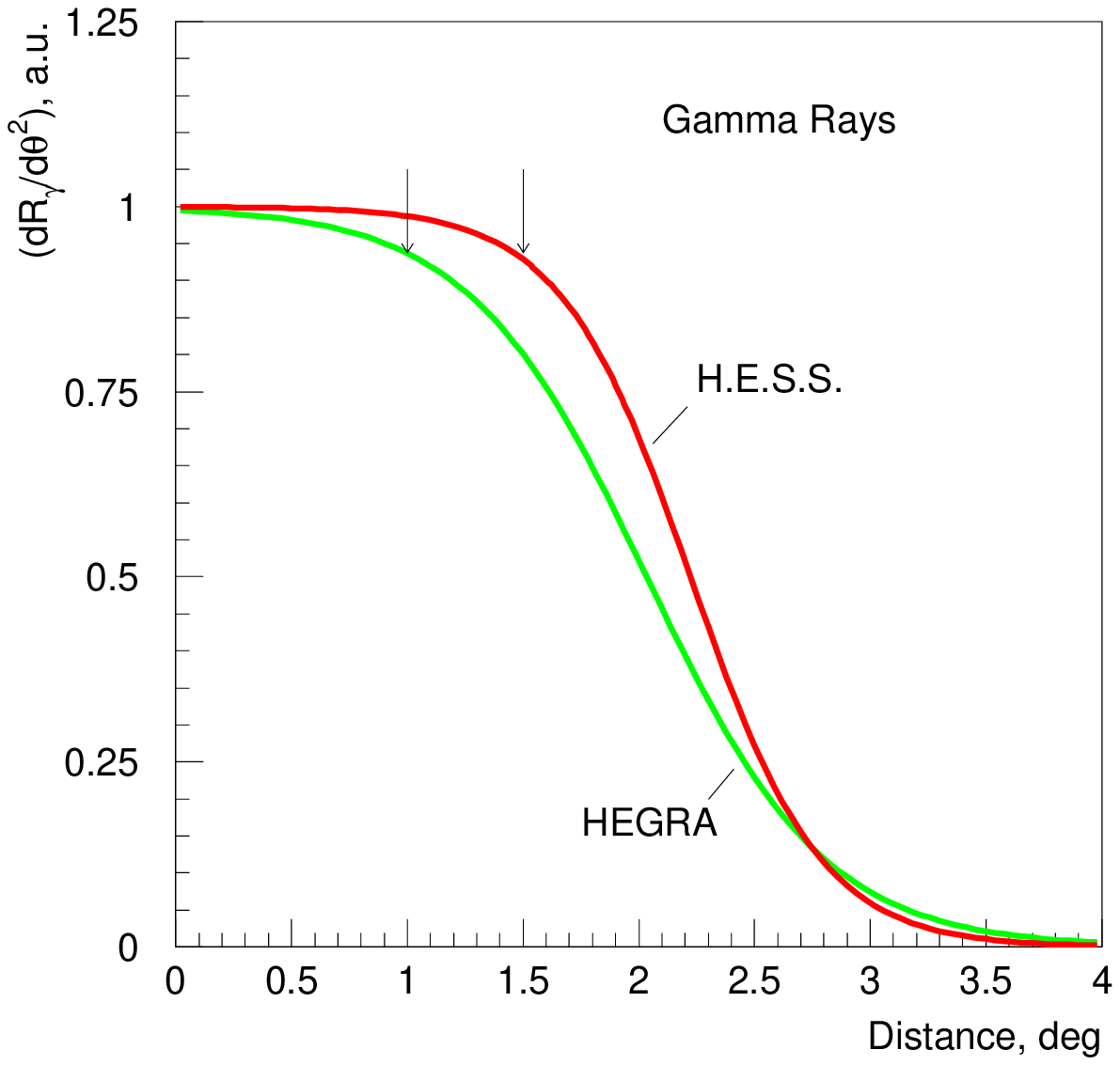}
\end{center}
\caption{Distribution of the error in the reconstructed shower core
  for H.E.S.S. (left panel). The efficiency of the OFF-axis
  $\gamma$-ray observations with the HEGRA \& H.E.S.S. (right panel).} 
\end{figure}
\begin{table}[h]
\caption{Accuracies in the stereo reconstruction.}
\begin{tabular}{lll}
                              & HEGRA & H.E.S.S.\\ \hline
$\bullet$ Angular resolution, [deg]  & $0.07^\circ$ & $0.1^\circ$ \\
$\bullet$ Source localization, arcsec & 40 & 10 \\
$\bullet$ Error in impact distance, [m] & 7 & 10 \\
$\bullet$ Energy resolution, $\rm \delta E/E$ & $12$\% & 20\% \\
$\bullet$ Error in shower maximum, [km] & 0.5 & $1$ \\
$\bullet$ Cosmic Ray rejection  & 0.01$\div$0.06 & 0.01$\div$0.1 \\
\end{tabular}
\end{table}
\noindent
one can calculate the corresponding rate for H.E.S.S. which could be about
150 $\gamma$-rays per min ! Such enormous statistics will allow
the measurements of the spectral shape for the Crab like $\gamma$-ray
sources after less then 1~hr of observations.   
The maximum $\gamma$-ray detection rate for H.E.S.S. corresponds to
the $\gamma$-ray energy within $\simeq$50-100~GeV (see Figure~1). 
The uncertainty 
in the position of the peak is caused by the current uncertainties of
the efficiencies along the over-all photon-to-photoelectron propagation
chain. One can break the entire dynamic energy range
(extending up to $\simeq 10$~TeV and $\simeq 50$~TeV for the
observations at high and low elevations, respectively) into two
parts (see Figure~1). As for the {\it high performance region} (above
100~GeV) H.E.S.S. will enable the high quality stereoscopic observations
with the ability of high precision spectroscopy based on the images of 
a large size as well as high trigger multiplicity. At such energies
the conventional data analysis developed for HEGRA could be of general
use. It is important to note that the effective collection area of
H.E.S.S. is larger then HEGRA, by a factor of $\simeq 5$, above 1~TeV.
It provides very high sensitivity of the H.E.S.S. in the multi-TeV
energy region. In the {\it sub-threshold region} (see Figure~1) the
\v{C}erenkov light images are substantially affected by the
telescope's trigger, have rather small angular size and poorly defined
shape due to the large fluctuations. In addition for these images the
night sky light contamination might be a severe problem. Nevertheless, 
the search for the periodic $\gamma$-ray signals
(e.g., from the Pulsars) in the energy range below 50~GeV could well be
physically expedient, eventhough it may need to develop a specific
analysis algorithms in particular for these observations. 

\section*{Stereo imaging}

For a system of IACTs the {\it stereoscopic} (geometrical) {\it 
reconstruction} of the air showers means the direct measurement of the 
orientation of the shower axis in space; determination of the position 
of the shower core in the observation plane; measurement of the
angular dimensions of a shower, and height of its maximum (the details
of such reconstruction could be found elsewhere). The major
parameters of the H.E.S.S. performance (phase I), comprising 4 IACTs
in a system, are summarized in Table~2.    

The important issue of the H.E.S.S. performance is a sensitivity to
the diffuse $\gamma$-rays over the field of view (FoV) (see
Figure~2). The analysis of the Monte Carlo events for H.E.S.S. doesn't 
show a significant decrease in the angular resolution for the OFF-axis 
events whereas the detection rate drops down rather fast beyond
$2^\circ$ (see Figure~2).

Although a numerous attempts have been undertaken in past in order to improve
the conventional stereoscopic algorithms, neither of those could gain
much. The HEGRA, with its ability of the individual event time sampling, did
not reveal a dominant use of the time profiles of the \v{C}erenkov light
flashes from the air shower (at least for the minimum time bins of
8~ns). 
On the contrary, applied after imaging analysis, the timing technique does 
not help at all. However, one could hope for a 0.5$\div$1~GHz signal
processing devices the advantage of the event time sampling might be
of great importance (e.g., for the night sky noise reduction etc). 

\section*{H.E.S.S. Sensitivity}

Provided with a 50~hrs observational time looking at a point source,
H.E.S.S. could see the $\gamma$-rays of the flux as low as 
\begin{equation}
\rm F^{p.s.}_\gamma(>100~GeV) \simeq 10^{-11} cm^{-2}s^{-1}
\end{equation}
at the 5$\sigma$ confidence level with the $\gamma$-ray statistics
exceeding 10 $\gamma$'s. The minimum detectable flux for an extended
$\gamma$-ray source was estimated as 
\begin{equation}
\rm F_\gamma{ext} \simeq (\Theta_0/0.1~deg)\cdot F^{p.s.}_\gamma
\end{equation}
No doubt such impressive sensitivity of future advanced ground based $\gamma$-ray
instrument allow to explore very effectively the sources of ``non-thermal emission in
the Universe''.  



\end{document}